\documentclass[twocolumn]{webofc}
\usepackage[varg]{txfonts}  
\usepackage{color}
\woctitle{ISVHECRI 2014}
\begin{document}
\title{CRPropa: a public framework to propagate UHECRs in the universe}

\author{R. Alves Batista\inst{1}\fnsep\thanks{\email{rafael.alves.batista@desy.de}} \and
        M. Erdmann \inst{2} \and C. Evoli \inst{1} \and K. -H. Kampert \inst{3} \and D. Kuempel \inst{2} \and G. M{\"u}ller \inst{2} \and G. Sigl \inst{1} \and A. van Vliet \inst{1} \and D. Walz \inst{2} \and T. Winchen \inst{3}
}

\institute{University of Hamburg, II. Institute for Theoretical Physics, Hamburg, Germany
\and
           RWTH Aachen University, Physikalisches Institut IIIa, Aachen, Germany
\and
          University of Wuppertal, Department of Physics, Wuppertal, Germany
          }

\abstract{

To answer the fundamental questions concerning the origin and nature of ultra-high energy cosmic rays (UHECRs), it is important to confront data with simulated astrophysical scenarios. These scenarios should include detailed information on particle interactions and astrophysical environments. To achieve this goal one should make use of computational tools to simulate the propagation of these particles. For this reason the CRPropa framework was developed. It allows the propagation of UHECRs with energies  $\gtrsim$10$^{17}$ eV and secondary gamma rays and neutrinos. The newest version, CRPropa 3, reflects an efficient redesign of the code as well as several new features such as time dependent propagation in three dimensions, galactic magnetic field effects and improved treatment of interactions, among other enhancements.

}

\maketitle

\section{Introduction}

The origin, nature and mechanisms of acceleration of ultra-high energy cosmic rays (UHECRs) are unanswered issues in astroparticle physics. To interpret the available experimental data one needs to model the source properties such as spectral index, mass composition and maximum acceleration energy, the distribution of sources, and the intervening cosmic magnetic fields. Effects arising from the interaction of cosmic rays with the pervasive photon backgrounds can also be relevant when constructing realistic scenarios.

Three observables measured in cosmic ray experiments are the spectrum, arrival directions and mass composition, the last one indirectly inferred from other observables, based on hadronic interaction models. They are affected by the interaction of UHECRs with photon fields, matter, as well as extragalactic and galactic magnetic fields. Any scenario aiming to elucidate the fundamental questions regarding the origin and nature of ultra-high energy radiation should explain these observables simultaneously, taking into account propagation effects such as energy losses and magnetic deflections. Therefore, the development of computational tools to simulate realistic scenarios that fit the data is necessary. For this reason the CRPropa code \cite{armengaud2007,kampert2013,alvesbatista2013,alvesbatista2014} was created.

This article is organized as follows: in section 2 we briefly review the photon backgrounds present in the universe and the interactions taking place at ultra-high energies; in section 3 we describe the structure of the code; section 4 contains short descriptions of the new features of CRPropa 3; in section 5 we present some applications; and in section 6 we present the concluding remarks and outlook.

\section{Interactions and photon backgrounds}

Ultra-high energy cosmic rays lose energy during their propagation to Earth mainly through four processes: pair production, pion production, photodisintegration (in the case of nuclei) and adiabatic expansion of the universe. 

The universe is permeated by photons with different wavelengths, which compose the extragalactic background light (EBL). The two main photon fields that should be taken into account in the propagation of UHECRs in the universe are the cosmic microwave background (CMB) and the cosmic infrared background (CIB). The first can be analytically estimated, whereas the second has to be obtained from observations. A third astrophysical background is the Universal Radio Background (URB). It has considerable effects only at $\gtrsim$10$^{22}$ eV, or in the development of electromagnetic cascades, which can be inhibited depending on the density of radio photons.

The redshift evolution of the number density of the cosmic microwave background is given by
\begin{equation}
 n_{CMB}(\epsilon,z)=\left(1+z\right)^2 n\left( \frac{\epsilon}{1+z},0\right),
\end{equation}
where $\epsilon$ is the energy of the background photon. For the CIB the number density is determined through observations. Its redshift evolution is not trivial, so that a complete modeling of this problem requires measurements of the density of photons at different redshifts.

The energy loss length for a nucleus of atomic number $Z$ and mass $A$ through production of electron/positron pairs can be written as \cite{kampert2013}
\begin{equation}
	\frac{dE}{dt}=3\alpha \sigma_T Z^2 h^{-3} (m_e c^2 k_B T)^2 f(\Gamma), 
\end{equation}
where $\alpha \approx 1/137$ is the fine structure constant, $\sigma_T$ is the Thomson cross section, $h$ the Planck constant, $k_B$ the Boltzmann constant, $\Gamma$ the Lorentz factor and $f(\Gamma)$ a function taken from ref. \cite{blumenthal1970}. The threshold energy for this interaction is $E_{thr} \approx \mathrm{5} (\mathrm{meV} / \epsilon ){\  }  \mathrm{EeV}$.

Photopion production occurs when an EBL photon is scattered by a nucleon. In the case of protons, the two main interaction channels are
\begin{equation*}
	p + \gamma \rightarrow \Delta^+ \rightarrow
	\begin{cases} 
		p + \pi^0 \\
		n + \pi^+
	\end{cases}.
\end{equation*}
The energy threshold for this process is $E_{thr} \approx \mathrm{70} (\mathrm{meV} / \epsilon) {\  } \mathrm{EeV}$. For a CMB photon with $\epsilon \approx \mathrm{0.6{\  }meV}$, $E_{thr} \approx \mathrm{4 \times 10^{19} {\  } eV}$, which is the expected energy for the well-known Greisen-Zatsepin-Kuzmin (GZK) cutoff.
The pions produced through the interaction between nucleons and EBL photons decay as follows:
\begin{equation*}
	\begin{cases}
	\pi^+ \rightarrow \mu^+ + \nu_\mu \\
	\pi^0 \rightarrow \gamma + \gamma
	\end{cases}.
\end{equation*}
This process is extremely important for multimessenger studies due to the production of secondary gamma rays and neutrinos. In CRPropa photopion production interactions are treated using the SOPHIA code \cite{muecke2000}. The mean free path for this interaction for nuclei can be written as a combination of the ones for protons and neutrons. 

The interaction of atomic nuclei with EBL photons causes these nuclei to split into parts, through a photodisintegration process. In CRPropa photonuclear cross sections are obtained from the TALYS code \cite{koning2008}. The mean free path for this process can be written in terms of the cross section $\sigma$ as follows:
\begin{equation}
\lambda^{-1}(\Gamma) = \frac{1}{2\Gamma^2} \int \limits^{\epsilon_{max}}_{\epsilon_{min}} \int\limits^{2\Gamma \epsilon}_0 n(\epsilon,z) \frac{1}{\epsilon^2}  \epsilon' \sigma(\epsilon') d\epsilon'd\epsilon,
\end{equation}
where $\Gamma$ is the Lorentz factor, $\epsilon$ the photon energy, $\sigma(\epsilon')$ the cross section of the nucleus-photon interaction, and $\epsilon_{max}$ the maximum energy of the background photon, which is $\sim$10 meV for the CMB and $\sim$100 eV for the CIB.

Unstable nuclei produced during photopion production or photodisintegration can have short lifetimes compared to the propagation length, and suffer decays during their trajectory. Our treatment of decays encompasses all relevant processes, namely $\alpha$, $\beta^+$ and $\beta^-$ decays, and proton and neutron drippings, with tabulated lifetimes from NuDat \footnote{For details refer to the website {\url{http://www.nndc.bnl.gov/nudat2/}}.}.

\begin{figure}[h!]
	\includegraphics[width=\columnwidth]{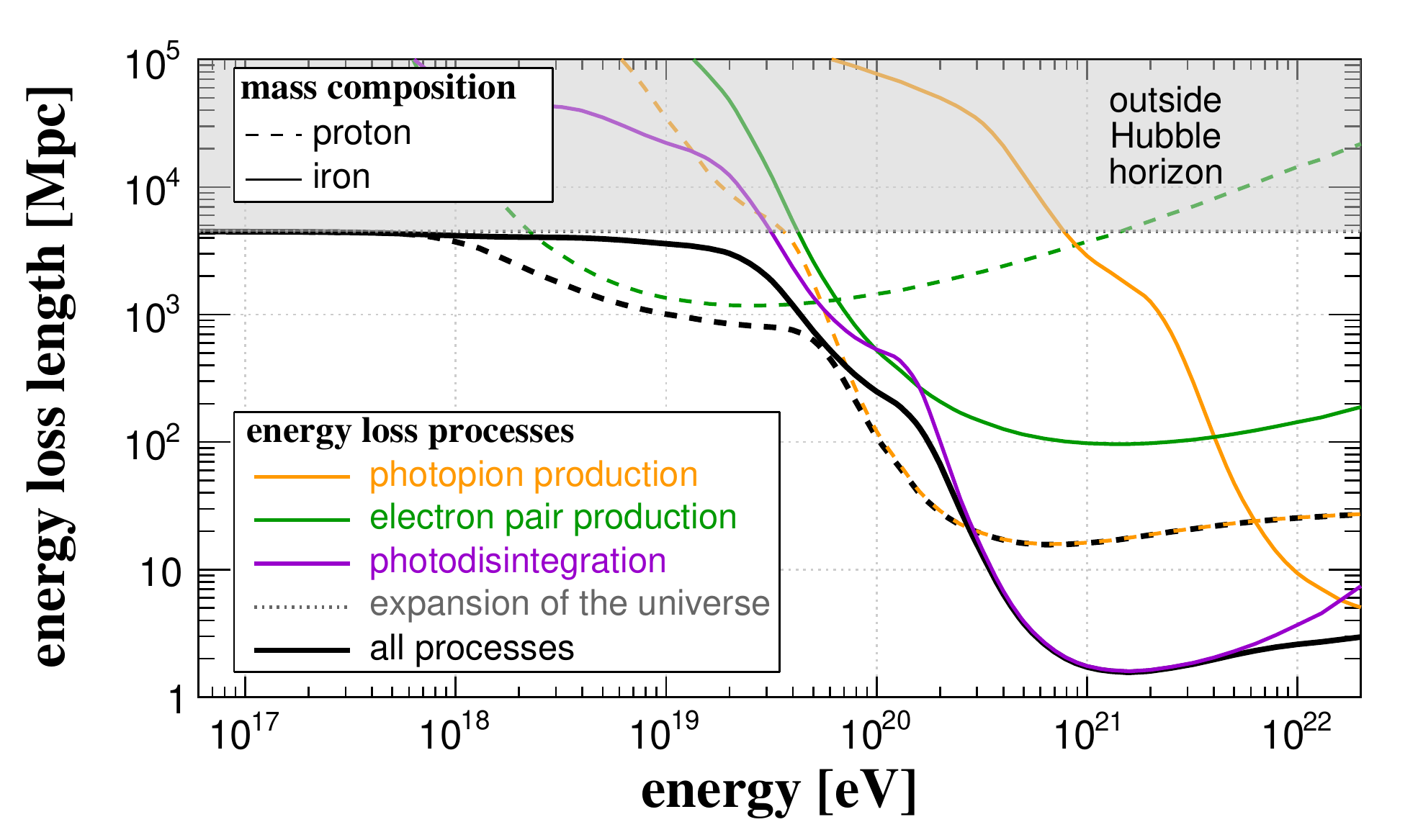}
	\caption{Energy loss lengths for different processes: photopion production (orange), electron pair production (green), photodisintegration (purple), adiabatic expansion of the universe (gray) and total (black). Solid lines are for iron nuclei, and dashed lines for protons. The photon backgrounds used here are the cosmic microwave background and infrared background from ref. \cite{kneiske2004}.}
	\label{fig:ell}
\end{figure}

The expansion of the universe itself is another source of energy loss. This adiabatic energy loss is given by
\begin{equation}
\label{eq:E_z}
 E=\frac{E_0}{1+z},
\end{equation}
where $E_0$ is the initial energy.

The energy loss length for photopion production, photodisintegration, pair production and adiabatic expansion of the universe are summarized in figure \ref{fig:ell}, for the case of iron and proton primaries.

\section{Code structure}

CRPropa 3 \cite{alvesbatista2013,alvesbatista2014} is a reformulation of the previous versions of the code \cite{kampert2013,armengaud2007}. It is written in C++ with Python bindings. It inherited all features from CRPropa 2 and added new functionalities. The modular structure of this new version allows the user to extend the code for other applications. The modularity comes from the construction of the code, which handles separately sources, observers, particles and each interaction. Individual independent modules alters the property of the `Candidate' class, which stores all details of the particle propagation, such as position, energy, type of particle, etc. These properties are updated at each step of propagation until a breaking condition is met or detection happens. CRPropa 3 supports shared memory parallel processing using OpenMP\footnote{{\url{www.openmp.org}}}, which allows fast simulations spanning a wide range of parameters to compare different scenarios.

\section{New features}

\subsubsection*{Four-dimensional propagation}

The simulation of UHECR propagation can be done in a one-dimensional (1D) environment, which allows the incorporation of cosmological effects such as the redshift dependence of the photon backgrounds, energy losses due to the adiabatic expansion of the universe, and source evolution, or in a three-dimensional (3D) environment, which can be used if one is interested in arrival directions, in addition to the spectrum and mass composition. In this case, cosmological effects cannot be taken into account because the information regarding the effective trajectory length of the particles, and therefore the redshift, is not known beforehand. The inclusion of cosmological effects in three-dimensional simulations can be done by tracking the particles not only in the three spatial coordinates, but also in time, within a four-dimensional (4D) approach. This feature is now included in CRPropa 3.

\subsubsection*{Galactic Magnetic Field}

CRPropa 3 allows the propagation of UHECRs considering several models of the galactic magnetic field (GMF). Implemented models include bissymmetric and antisymmetric spirals, as well as a toroidal field. The recent model by Jansson \& Farrar \cite{jansson2012a, jansson2012b} is also implemented, including regular, striated, and random turbulent components. 

Any 3D or 4D simulation of the extragalactic propagation of UHECRs can be corrected for the effects of the galactic magnetic field. This is done {\it a posteriori} through a lensing technique first implemented in the PARSEC code \cite{bretz2014}, and adapted for CRPropa. For each energy there is a lens (matrix) which maps the directions of cosmic rays arriving at the edge of the galaxy to a direction observed at Earth. In this case no energy losses are considered.

\subsubsection*{Extragalactic Magnetic Fields} 

Realistic scenarios of source distribution and magnetic fields are important to constrain models of UHECRs. In this context, cosmological magnetohydrodynamical (MHD) simulations of the local universe can provide information regarding the matter distribution and intervening magnetic fields. In previous versions of CRPropa it was possible to consider a simple turbulent magnetic field, as well as magnetic fields from MHD simulations through uniformly spaced grids. The usage of multiresolution grids in MHD simulations has increased recently, and these higher resolution non uniform grids could be useful for cosmic ray propagation. For this reason CRPropa 3 interfaces with external codes to handle MHD simulations from the SPH (Smooth Particle Hydrodynamics) code Gadget \cite{dolag2009} and AMR (Adaptative Mesh Refinement) code RAMSES \cite{teyssier2002}.

\subsubsection*{UHE photon propagation with the EleCa code}

Ultra-high energy photons can be primary particles, emitted by an astrophysical object, or secondaries, generated through the interaction of primary cosmic rays with photon backgrounds. The interaction of these photons with the EBL induces the development of an electromagnetic cascade, which can be propagated within CRPropa with the external codes DINT \cite{lee1998}, which calculates the photon spectrum by solving kinetic equations, or EleCa \cite{settimo2015}, a Monte Carlo code for propagating UHE photons in the universe. 

The resulting photon spectrum calculated by DINT ranges from 10$^8$ eV up to 10$^{23}$ eV, while EleCa is restricted to higher energies ($\gtrsim$ 10$^{16}$ eV). Because DINT is based on transport equations, it is more efficient for lower energies, allowing a faster calculation of the spectrum compared to the more precise particle-by-particle Monte Carlo approach of EleCa. For now the propagation of photons with EleCa is done only in one dimension, and the effects of magnetic fields on the cascades are taken into account using a small angle approximation.

\subsubsection*{Updated photodisintegration cross sections}

As mentioned before, in CRPropa 3 the treatment of photodisintegration is done using tabulated values for the cross section, taken from the TALYS code \cite{koning2008}. The previous version, CRPropa 2, uses photodisintegration cross sections from TALYS 1.0, whereas CRPropa 3 includes the up-to-date TALYS 1.6\footnote{For detailed information on the changes made between TALYS versions 1.0 and 1.6, refer to the documentation in the website {\url{www.talys.eu}}. } cross sections for A$\geq$12, which improves the binning of the data around resonances, extends the available energy range, and enhances some reactions, among other improvements.

\begin{figure}[h!]
	\includegraphics[width=\columnwidth]{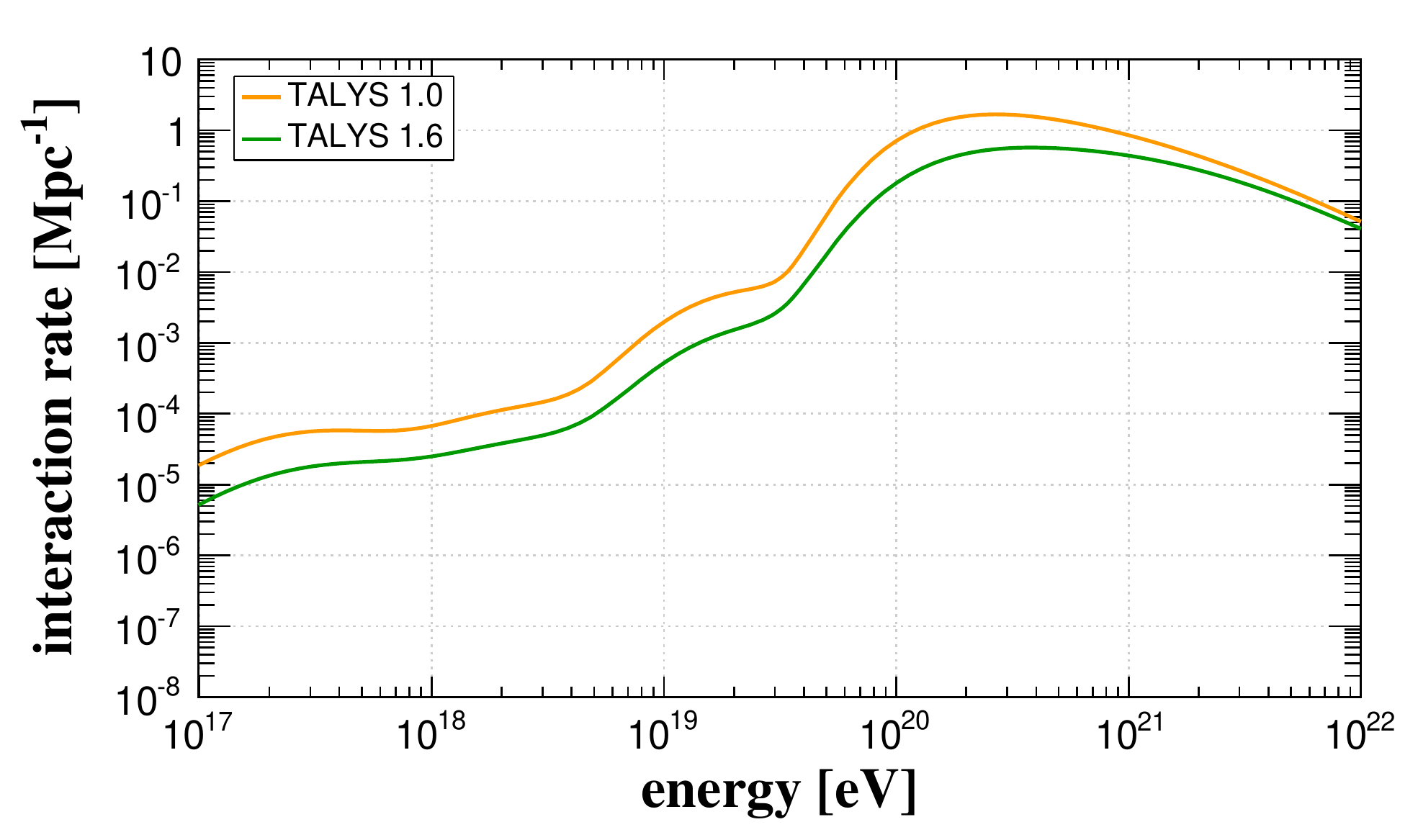}
	\includegraphics[width=\columnwidth]{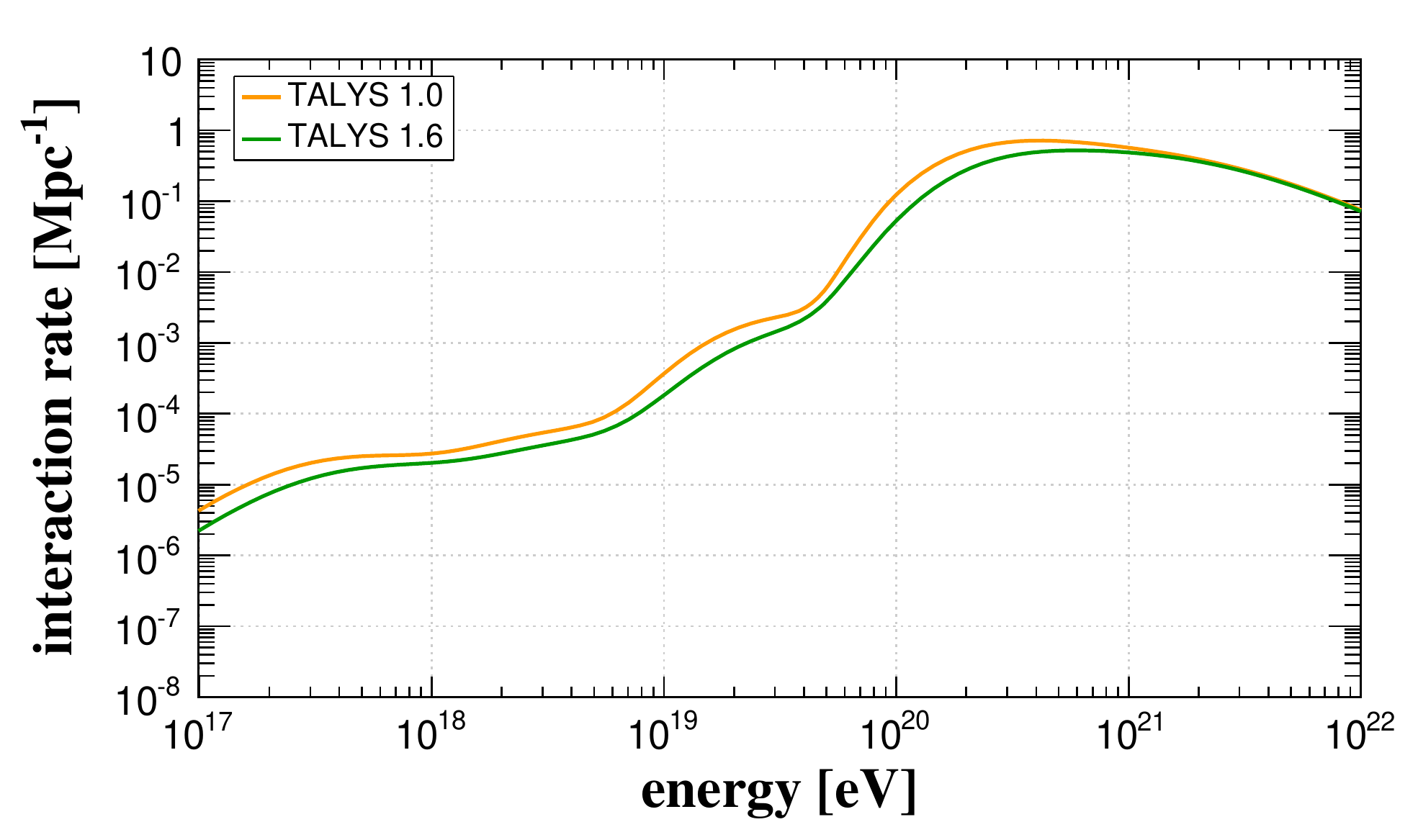}	
	\caption{Photodisintegration interaction rates for $^{12}_{6}$C (top) and $^{16}_{8}$O (bottom) for TALYS 1.6 (green) and TALYS 1.0 (orange).}
	\label{fig:pdir}
\end{figure}

The impact of the new photodisintegration cross sections provided by TALYS 1.6  on the photodisintegration rates are approximately the same as the ones from TALYS 1.0 for the lightest and heaviest nuclei, but they differ significantly for intermediate masses. In figure \ref{fig:pdir} the impact of these differences on the interaction rate of UHECRs with EBL photons are shown for the case of carbon-12 and oxygen-16. 

\subsubsection*{Cosmic Infrared Background models}

CRPropa 3 provides several options of models of the CIB. The user can choose between a set of predefined CIB models, namely the ones by Kneiske {\it et al.} \cite{kneiske2004}, Franceschini \cite{franceschini2008}, Dole {\it et al.} \cite{dole2006}, Kneiske \& Dole lower limit \cite{kneiske2010} and Stecker \cite{stecker2012}.

\section{Applications}

\begin{figure}[h!]
	\includegraphics[width=\columnwidth]{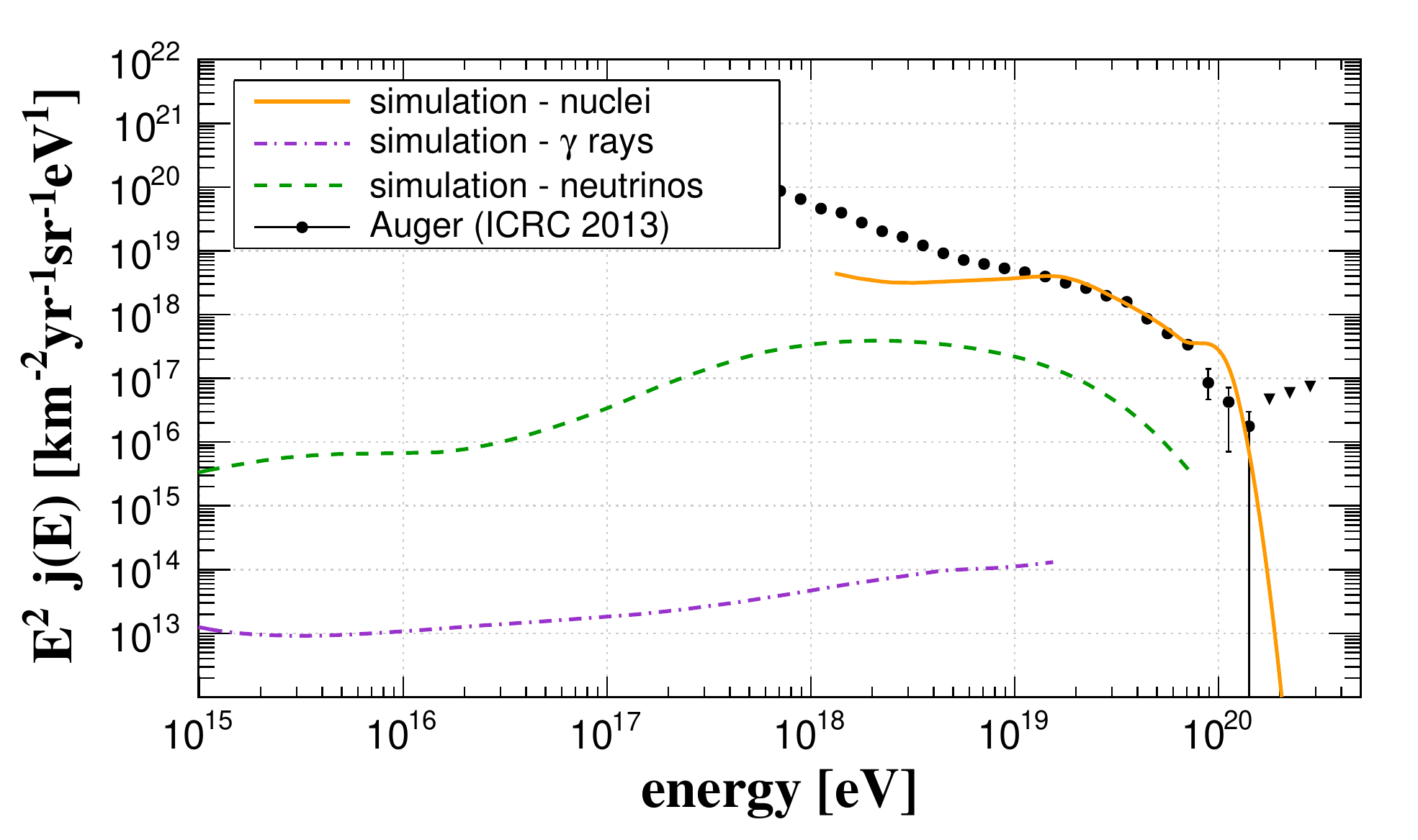}
	\caption{Simulated UHECR (solid line) and secondary gamma ray (dot-dashed line) and neutrino (dashed line) spectra. The markers correspond to the Auger 2013 spectrum \cite{auger2014}.}
	\label{fig:example1D}
\end{figure}

First we illustrate an application of CRPropa within a multimessenger approach, propagating UHECRs and obtaining the photon and neutrino counterparts. We assume a uniform distribution of sources with spectrum $dN/dE \propto E^{-2}$. The infrared background model assumed is the one by Kneiske {\it et al.} \cite{kneiske2004}. The secondary photons and neutrinos spectra produced from the interaction of protons with background photons were also computed. The spectra for cosmic rays and secondaries are shown in figure \ref{fig:example1D}. In this figure data from the Pierre Auger Observatory \cite{auger2014} are displayed for the sake of comparison. 

We now present the results of a three-dimensional simulation. We use the large scale matter distribution from Dolag {\it et al.} \cite{dolag2005}, constrained in such a way to reproduce the observational data from cosmological surveys. This box, henceforth called matter distribution grid, is a cube grid of approximately 132 Mpc with a spacing of $\sim$300 kpc.
Instead of using the magnetic field distribution from this MHD simulation, we use the one from Miniati \cite{miniati2002}, also used in several subsequent works \cite{sigl2003,armengaud2005} for the propagation of UHECRs. This model has a higher magnetic field strength compared to Dolag {\it et al.}. We use the profile magnetic field-density distributions from the Miniati simulation to obtain a relation between magnetic field and density. We then create a modulation grid by replicating the matter distribution grid, replacing the density in each cell by the corresponding magnetic field strength from the profile. The modulation grid has 256$^3$ cells covering a volume of approximately (132 Mpc)$^3$. It is used to modulate another 256$^3$ grid with volume $\sim$(13.2 Mpc)$^3$ containing a realization of a turbulent Kolmogorov field with coherence length 500 kpc, periodically repeated to cover the complete simulation volume.

To illustrate the propagation including effects of magnetic fields and matter distribution we arbitrarily assume a scenario  composed of four species of atomic nuclei, hydrogen, helium, nitrogen and iron, with fraction 1, 0.5, 0.25 and 0.125, respectively. The sources follow the large scale distribution from Dolag {\it et al.} \cite{dolag2005}, and emit particles with a differential spectrum proportional to $E^{-1.8}$.  The maximum rigidity of the source is $R_{max} = \mathrm{10^{19.8}}$ eV. The maximum trajectory length considered was 2 Gpc.

\begin{figure}[h!]
	\includegraphics[width=\columnwidth]{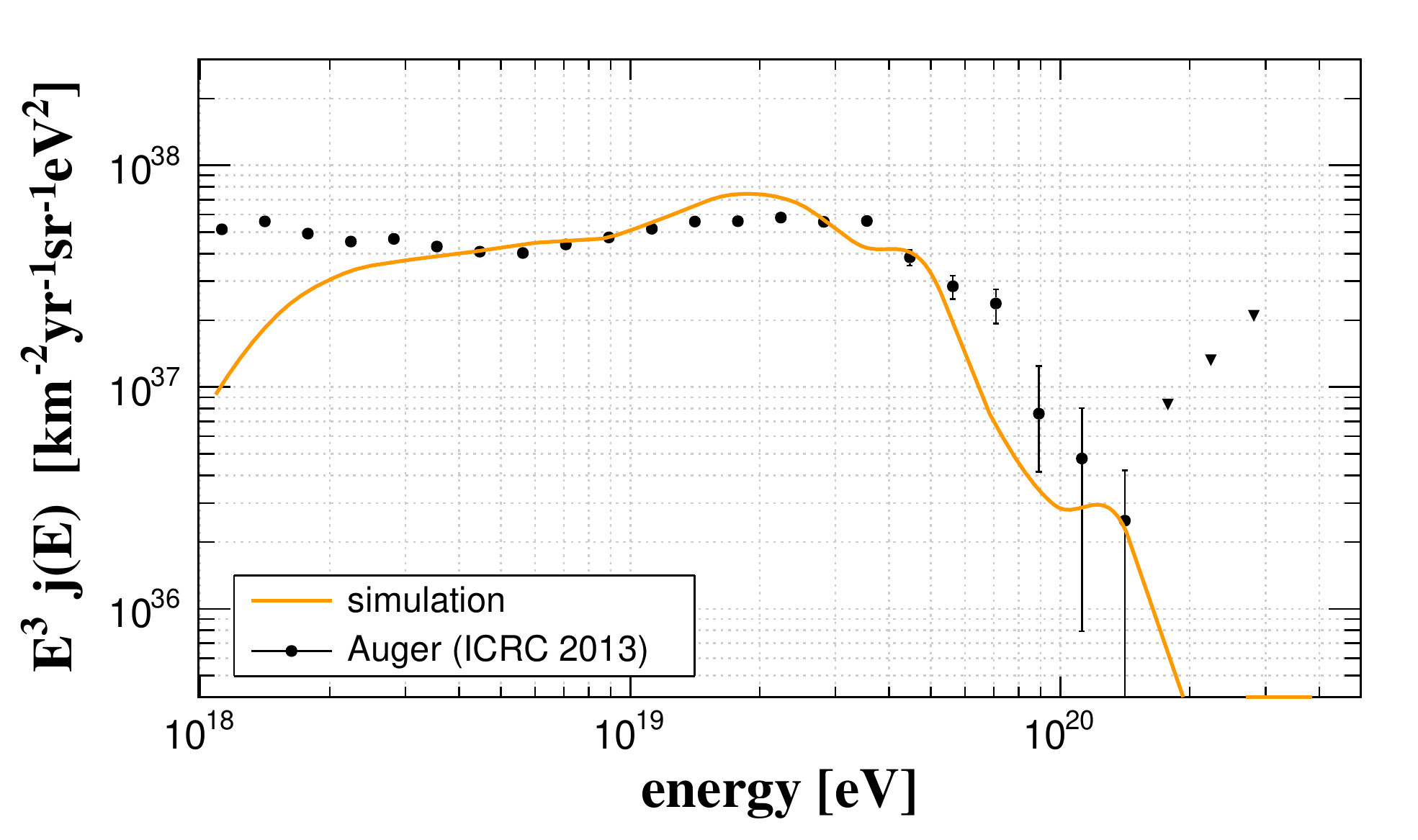}
	\caption{UHECR spectrum for the simulated scenario. The line correspond to the simulated scenario, and the markers to measurements from the Pierre Auger Observatory \cite{auger2014}.}
	\label{fig:3Dexample-spec}
\end{figure}

In figure \ref{fig:3Dexample-spec} the cosmic ray spectrum above 10$^{18}$ eV is shown. 

\begin{figure}[h!]
	\includegraphics[width=\columnwidth]{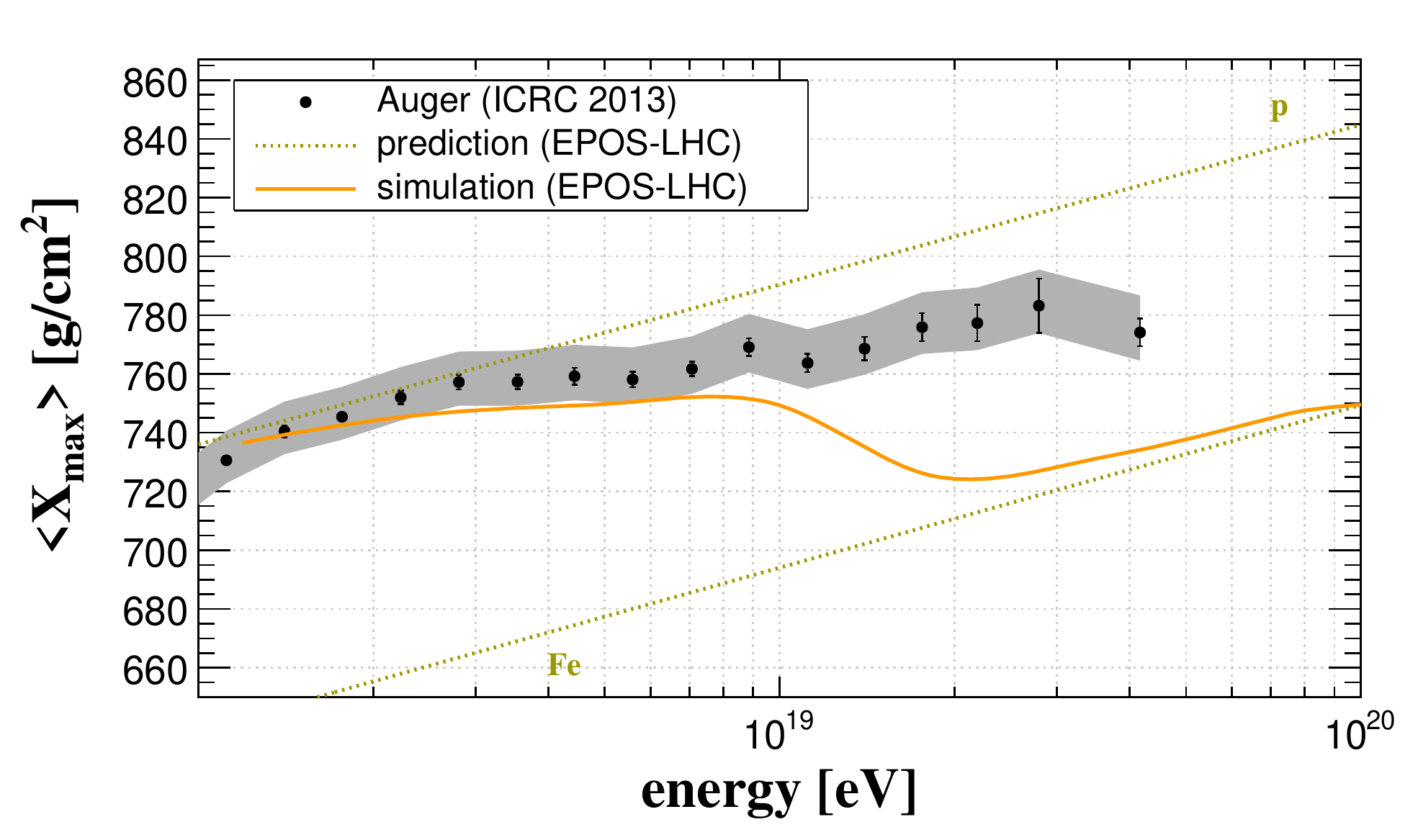}
	\includegraphics[width=\columnwidth]{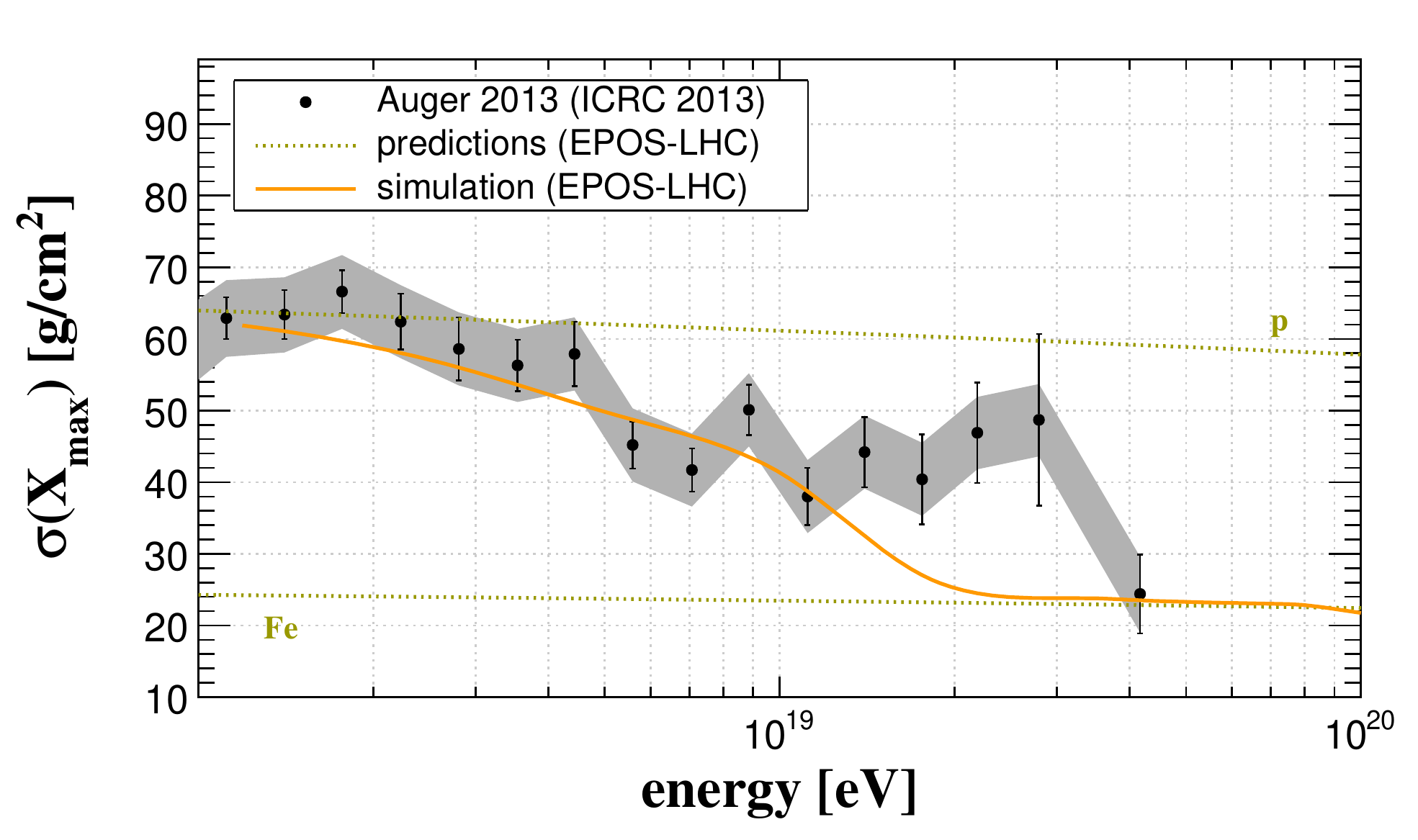}
	\caption{Estimated $X_{max}$ and $\sigma(X_{max})$ for the simulated data (orange line). Black markers correspond to data measured by the Pierre Auger Observatory \cite{auger2014}, and gray regions are the systematics. Dotted lines are the predicted composition for pure proton and pure iron scenarios, according to the EPOS-LHC hadronic interaction model.}
	\label{fig:3Dexample-comp}
\end{figure}

The mass composition of the observed particles is expected to be different from the injected one, due to photodisintegration. In terms of experimental observables, a change in composition implies a change in the  depth of the shower maximum ($\langle X_{max} \rangle$). In figure \ref{fig:3Dexample-comp} the values of $\langle X_{max} \rangle$ and  $\sigma(X_{max})$ for this scenario are shown. They were estimated using the parametrization from refs. \cite{auger2013,auger2013b}, assuming the EPOS-LHC hadronic interaction model \cite{pierog2013}.

\begin{figure}[h!]
	\includegraphics[width=\columnwidth]{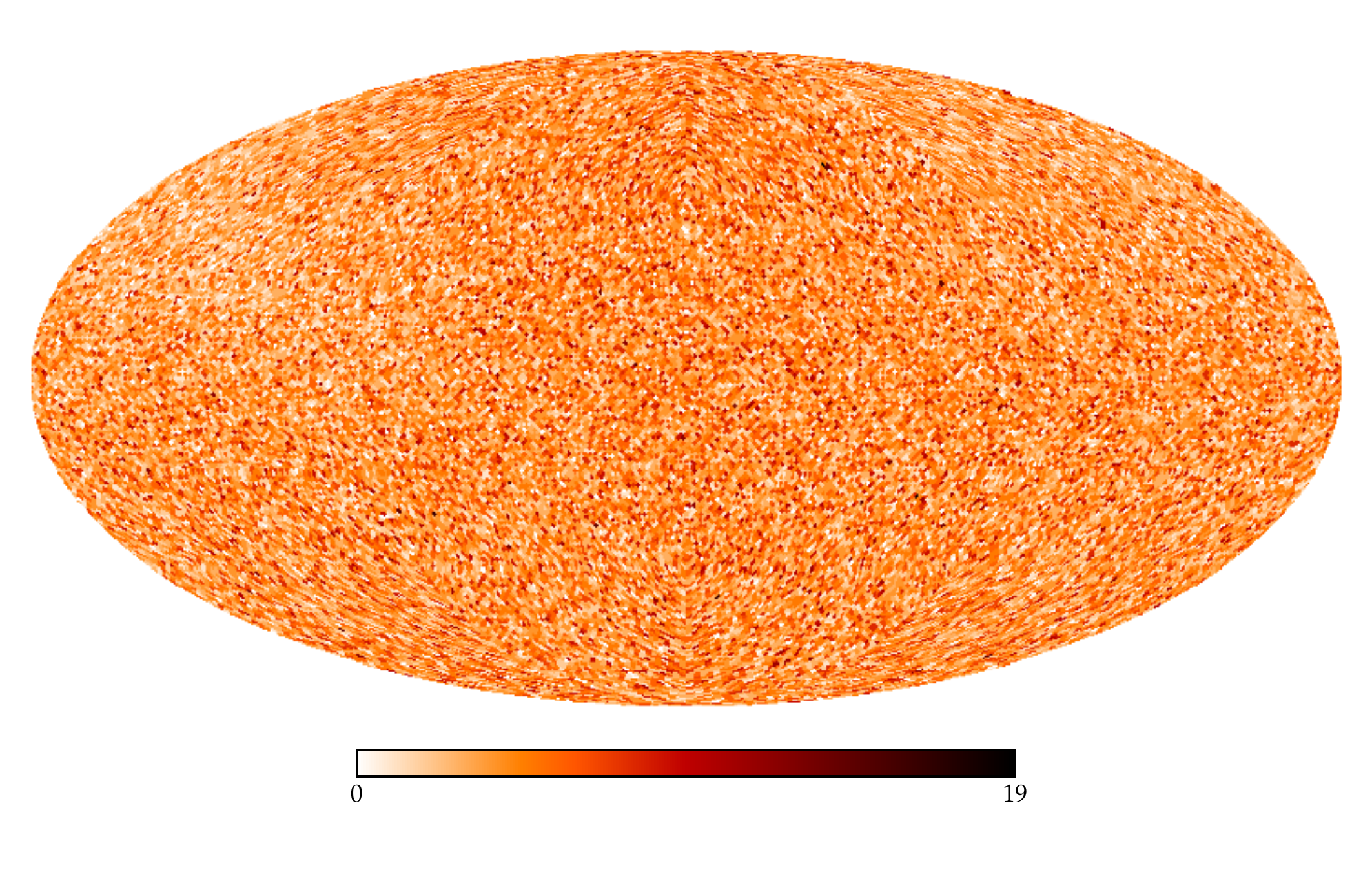}
	\includegraphics[width=\columnwidth]{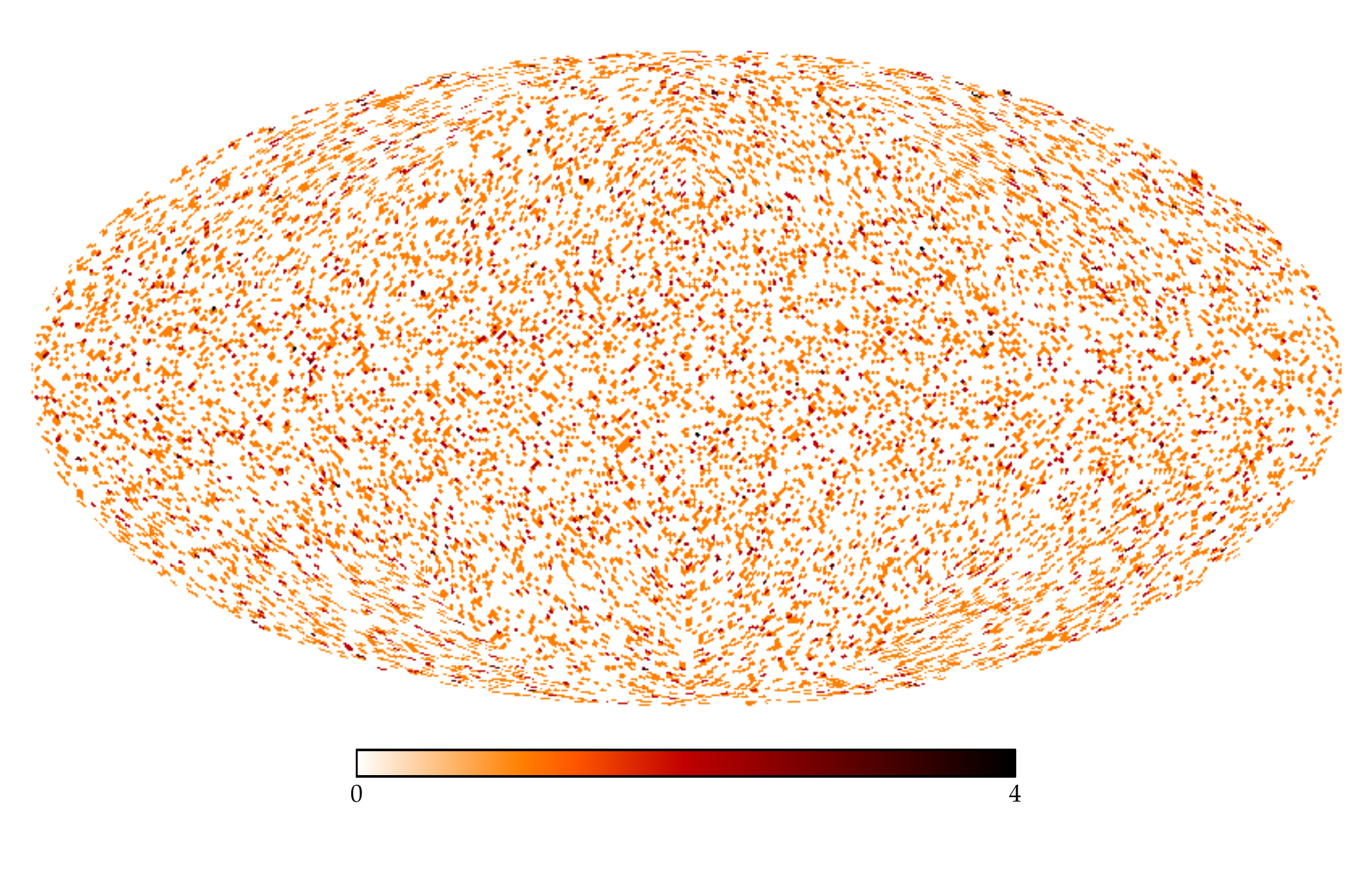}
	\caption{Skymaps for the simulated scenario without (top) and with (bottom) the effects of the galactic magnetic field. The color scale indicates the number of events per pixel.}
	\label{fig:3Dexample-sky}
\end{figure}

The skymaps containing the arrival directions of the simulated events are shown in figure \ref{fig:3Dexample-sky} considering only extragalactic deflections and including effects of the galactic magnetic field according to the model by Jansson \& Farrar \cite{jansson2012a,jansson2012b}.

\section{Summary and outlook}

The comparison between experimental data and theoretical models is crucial to understand the origin and nature of the UHECRs. Therefore, the development of computational tools that allow the detailed simulation of astrophysical scenarios including all relevant particle physics and astrophysical ingredients are essential to build realistic scenarios.

We have presented CRPropa 3, a public framework to propagate UHECRs,and secondary gamma rays and neutrinos in the universe. The code allows parallel processing, python steering and the inclusion of custom modules. The main new features are the four-dimensional propagation, galactic magnetic fields, new extragalactic magnetic field techniques, improved interaction tables and new CIB models. We have presented two applications for illustration purposes, and compared the results with measurements from the Pierre Auger Observatory. 

CRPropa 3 is in the final stages of development, is publicly available, and will soon be released. More information can be found on {\url{https://crpropa.desy.de}}.

\section*{Ackowledgements}

At the University of Hamburg this work was supported by the Deutsche Forschungsgemeinschaft (DFG) through the Collaborative Research Centre SFB 676 ``Particles, Strings and the Early Universe''  and by BMBF under grants 05A11GU1 and 05A14GU1. RAB acknowledges the support from the Forschungs- und Wissenschaftsstiftung Hamburg through the program ``Astroparticle Physics with Multiple Messengers''.
We also acknowledge support from the Helmholtz Alliance for Astroparticle Physics (HAP) funded by the Initiative and Networking Fund of the Helmholtz Association.

\bibliography{references}

\end{document}